\documentclass[12pt]{article}
\usepackage{amsmath}
\usepackage{amsfonts}
\usepackage{amssymb}
\usepackage{textcomp}
\usepackage[dvips]{graphicx}
\usepackage{epsfig}
\usepackage{bm}
\usepackage{dcolumn}
\usepackage{amsfonts}
\usepackage{color}
\usepackage[left=2cm,right=2cm,top=2cm,bottom=2cm]{geometry}
\usepackage[left=2cm,right=2cm,top=2cm,bottom=2cm]{geometry}

\begin{document}

\title{\textbf{A note on the arrow of time in nonminimally coupled scalar field FRW cosmology}}

\author{L.A. Le\'{o}n Andonayre$^{1}$\footnote{lal10@hi.is}, 
M.A. Skugoreva$^{2}$\footnote{masha-sk@mail.ru}, \\
A.V. Toporensky$^{2,3}$\footnote{atopor@rambler.ru},
T. Vargas$^{4}$\footnote{teofilo.vargas@gmail.com}\vspace*{3mm} \\
\small $^{1}$Science Institute, University of Iceland,
\small Dunhaga 3, 107 Reykjavík, Iceland\\
\small $^{2}$Kazan Federal University, Kremlevskaya 18, Kazan, 420008, Russia\\
\small $^{3}$Sternberg Astronomical Institute, Lomonosov Moscow State University,\\
\small Moscow, 119991, Russia\\
\small $^{4}$Grupo de Astronom\'{i}a SPACE and Grupo de F\'{i}sica Te\'{o}rica GFT, Facultad de Ciencias F\'{i}sicas, \\
\small Universidad Nacional Mayor de San Marcos, Lima-Peru}

\date{ \ }

\maketitle

\begin{abstract}

    We revisit the cyclic Universe scenario in scalar field FRW cosmology and check its applicability for a nonminimally coupled scalar field. We show that for the most popular case of a quartic potential and the standard nonminimal coupling this scenario does work. On the other hand, we identify certain cases where cyclic model fails to work and present the corresponding reasons for this.
\end{abstract}

\section{Introduction}
~~~~The question of time arrow in a scalar field Friedmann-Robertson-Walker (FRW) cosmology is rather intriguing since the corresponding cosmological equations of motion are dissipationless, and, then, formally admit a time reversal. Nevertheless, it is known that in some cyclic models based on a minimally coupled scalar field there are infinitely growing circles of the scale factor appearing for a ``typical'' initial conditions. Later this fact was linked to the time asymmetry of the effective equation of state of the scalar field. The fact that the effective equation of state for a minimally coupled scalar field is drastically different for the expansion and contraction stages has been shown in \cite{foster}, in particular, a power-law potential results in the $\omega_\mathrm{eff}=1$ asymptotic during contraction while being near $-1$ at expansion. Growing cycles in a cyclic Universe with a scalar field have been considered numerically in~\cite{kanekar2001recycling}, the connection of the growth rate of the scale factor with the time asymmetry of the effective equation of state has been studied in~\cite{sahni2012cosmological}. 

    We should, however, remind the reader that creating a cycling cosmological evolution itself (apart from the question of a preferred time direction) needs a special effort. A transition from expansion to contraction (a turn-around point) can be achieved within General Relativity (GR) by adding a small negative cosmological constant or by considering a positively curved Universe. As for a transition from contraction to expansion (a bounce point), it needs going beyond GR, and even in this case it is not easy. There are some models containing the necessary type of bounce, for example, in Loop Quantum Cosmology~\cite{PhysRevD.69.104008,dupuy2020hysteresis}, in a brane model with a timelike compact dimension~\cite{SHTANOV20031,Piao1,Piao2} or in models with negative energy (``ghost'') matter~\cite{Barrow}, but they are considered nowadays as rather exotic models. Therefore an artificial bounce put ``by hand'' is sometimes used to model the cycles. This means that when the energy density of the scalar field exceeds some value (usually assumed to be of the order of the Planck energy density), the Hubble parameter $H$ is set to change its sign.~\footnote{A transition from contraction to expansion is possible in GR cosmology driven by a minimally coupled scalar field if the spatial curvature is positive, see~\cite{Starobinsky1}. We should add that the GR bounce present in positively curved models is not suitable for our goals here since it requires the same effective equation of state for the scalar field at the contraction stage ($\omega_\mathrm{eff}$ should be close to $-1$) as for the expanding stage, so the condition for a time arrow to appear is not realized. Indeed, the dynamics for shallow scalar field potentials (when the bounce is a typical outcome for the contraction stage) shows the existence of quasiperiodic oscillations~\cite{TT,Matsui}, and no time arrow appears.}

    There are two possibilities for such a bounce considered earlier. Either we change only $H$, leaving the velocity of the scalar field $\dot\phi$ unchanged, or the sign of the velocity is reversed as well. In the former case we have a growing size of the Universe from cycle to cycle for the most of initial condition~\cite{sahni2012cosmological}. This model has been recently intensively studied not only from the viewpoint of FRW dynamics, but also including its perturbations (see, e.g.,~\cite{Piao3, Piao4}). As for the second option, the sign reversal for both $H$ and $\dot\phi$ in fact results in changing the time direction for the same trajectory, so that the Universe follows the same trajectory, and strictly speaking there should be no room for time arrow effects. However, they can appear even for this setting if small deviations from exact relations $H \to -H$ and $\dot \phi \to -\dot\phi$ are introduced~\cite{sahni2015arrow}. Such fluctuations near the bounce are natural to expect if the bounce is thought to occur due to some still unknown quantum effects. 

    All previous considerations have been made for a minimally coupled scalar field, where the abovementioned difference in the effective equation of state reaches its maximal for this type of the scalar field. Namely, it varies from about $\omega_\mathrm{eff}=-1$ at expansion to $\omega_\mathrm{eff}=+1$ at contraction. The goal of the present paper is to consider the possibility of a cosmological time arrow for nonminimally coupled scalar fields. We shall use only the first option and assume that after reaching some energy density the sign of $H$ reverses while the sign of $\dot\phi$ does not.

    Concluding the Introduction, we would like to point out that the scenario we consider here is different from the known ``Weyl curvature hypothesis'' which represents an alternative approach to the effective arrow of time problem. That approach has been put forward by R.~Penrose~\cite{Penrose79} who states that ``In terms of space-time curvature, the absence of clumping corresponds, very roughly, to the absence of Weyl conformal curvature (since the absence of clumping implies spatial isotropy and hence no gravitational principal null directions). When clumping takes place, each clump is surrounded by a region of nonzero Weyl curvature. As the clumping gets more pronounced owing to gravitational contraction, new regions of empty space appear with Weyl curvature of greatly increased magnitude. Finally, when gravitational collapse takes place and a black hole forms, the Weyl curvature in the interior region is still larger and diverges to infinity at the singularity.'' Consequently, the initial minimum gravitational entropy state is, according to Penrose, proportional to a maximum matter entropy state in which the initial matter is almost uniform and in a state of thermal equilibrium (see also~\cite{Gron05,Barrow02}). This proposal met some degree of success with a slight modification, and the early work on this conjecture was systematically summarized and commented on by Clifton et al. in~\cite{Clifton13}, where the authors also proposed a measure of gravitational entropy based on the square-root of the Bel-Robinson tensor of free gravitational fields, which is the unique totally symmetric traceless tensor constructed from the Weyl tensor. The results of our approach demonstrate the existence of a cosmological situation where the effective arrow of time is {\it not} associated with the Weyl tensor. This is so because this tensor vanishes identically for the FRW metric which we consider in the present paper.

\section{Nonminimal coupling}	
~~~~The Einstein-Hilbert action of the scalar-tensor model where a scalar field couples nonminimally to gravity has the following general form:
\begin{equation}
\label{S1}
S =\int d^4x\sqrt{-g}\left[\frac{1}{2}\left(\frac{1}{K}-\xi B(\phi)\right)R-\frac{1}{2}g^{\mu\nu}\partial_{\mu}\phi\partial_{\nu}\phi- V(\phi)\right ]\,,
\end{equation}
where $g=\mathrm{det}(g_{\mu\nu})$ is the determinant of the metric tensor, $R$ is the Ricci scalar, $V(\phi)$ the scalar field potential, $B(\phi)$ the nonminimal coupling function of the scalar field $\phi$, the constant $\xi$ controls the coupling between the scalar field and the Ricci scalar curvature, $K=8\pi G$, and we use $c=1$ and the metric signature $(-,+,+,+)$.  

    We assume a homogeneous distribution for the scalar field in the spatially flat FRW universe with the metric $ds^2=-dt^2+a^2(t)dl^2$, where $a(t)$ is the scalar factor. The time-time component of the Einstein equations and the energy-momentum tensor give the energy constraint equation
\begin{equation}
\label{00}
3 H^2\Big{(}1-K \xi B(\phi)\Big{)} = K \left(\frac{{\dot\phi}^2}{2} +V(\phi)+3\xi H \dot\phi B'(\phi)\right)\,,
\end{equation}
whereas the spatial diagonal components are 
\begin{equation}
\begin{array}{l}
\label{11}
(2\dot H+3H^2)\Big{(}1 - K\xi B(\phi)\Big{)}=\\
=K \left[-\frac{1}{2}{\dot\phi}^2+V(\phi)+\xi\Big{(}2H\dot\phi B'(\phi)+ {\dot\phi}^2 B''(\phi)+\ddot\phi B'(\phi)\Big{)}\right]\,,
\end{array}
\end{equation}
where differentiation with respect to time $t$ is denoted by a dot, the prime indicates a derivative with respect to the scalar field $\phi$, and $H \equiv {\dot a(t)}/a(t)$ is the Hubble expansion rate. 

    Equations (\ref{00}) and (\ref{11}) can be rewritten as
\begin{equation}
\label{00rho}
3 H^2= K\rho_{\phi}\,,
\end{equation}
\begin{equation}
\label{11p}
2\dot H+3H^2=-K p_{\phi}\,,
\end{equation}
where the scalar field energy density $\rho_{\phi}$ and pressure $p_{\phi}$ are~\cite{Polarski}
\begin{equation}
\begin{array}{l}
\label{rhopphi}
\rho_{\phi}=\frac{1}{2}{\dot\phi}^2+V(\phi)+3\xi\Big{(} H\dot\phi B'(\phi)+H^2 B(\phi)\Big{)},\\
p_{\phi}=\frac{1}{2}{\dot\phi}^2-V(\phi)-\xi\Big{(}2H\dot\phi B'(\phi)+ {\dot\phi}^2 B''(\phi)+\ddot\phi B'(\phi)+(2\dot H+3H^{2}) B(\phi)\Big{)}.
\end{array}
\end{equation}

    Variation of the action (\ref{S1}) with respect to the field $\phi$ results in the Klein-Gordon equation  
\begin{equation}
\label{KG}
\ddot\phi +3 H\dot\phi+\frac{1}{2}\xi R B'(\phi)+V'(\phi)=0\,
\end{equation}
and the Ricci scalar in flat FRW metric is given by $R=6(2H^2+\dot H)$. In what follows we consider only the standard form of nonminimal coupling where $B(\phi)=\phi^2$, the coupling constant $\xi<0$~\footnote{A positive $\xi$ for large $\phi$ would formally lead to antigravity domain with negative effective Newton constant. FRW dynamics allows penetrating into it, but a small anisotropy leads to a singularity at zero Newton constant~\cite{Starobinsky2}.} and the potential in the following form: $V=V_0\phi^n+\Lambda$, where $V_0>0$, $n>0$, $\Lambda<0$ are parameters.~\footnote{We consider here the nonminimal coupling models with a positive potential. In the case of a negative potential, a number of bouncing FRW solutions were found in~\cite{Starobinsky3,Starobinsky4,we1,Vernov} and in the more general anisotropic Bianchi I case in~\cite{Starobinsky5}. However, all these models require a positive $\xi$ and enters into an antigravity domain either before or after the bounce, so they are not suitable for a cycling scenario.}
\begin{figure}[!ht]
\centering
\includegraphics[scale=0.68]{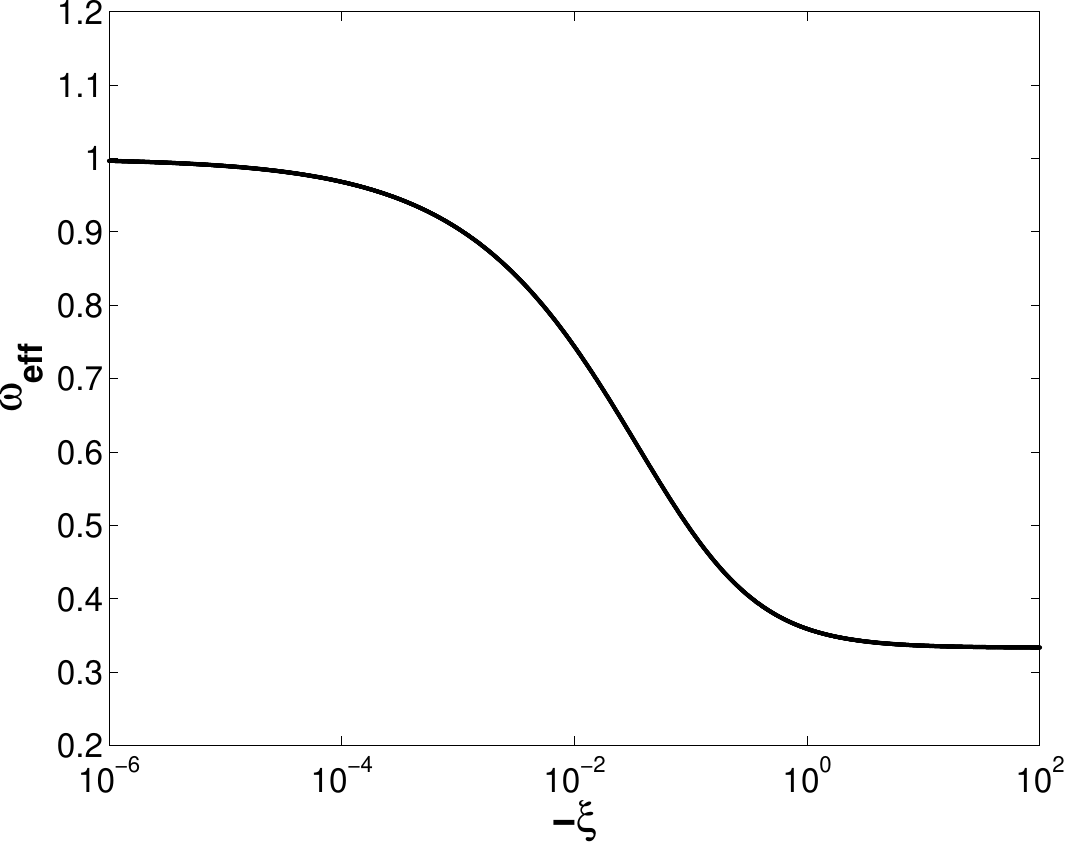}
\caption{The dependence of the effective equation of state parameter $\omega_\mathrm{eff}$ on the coupling constant~$-\xi$ in~(\ref{weff}) for the power-law asymptotic regime corresponding to an unstable node.}
\label{fig:weffvsxi}
\end{figure}
\begin{figure}[!ht]
\centering
\includegraphics[scale=0.68]{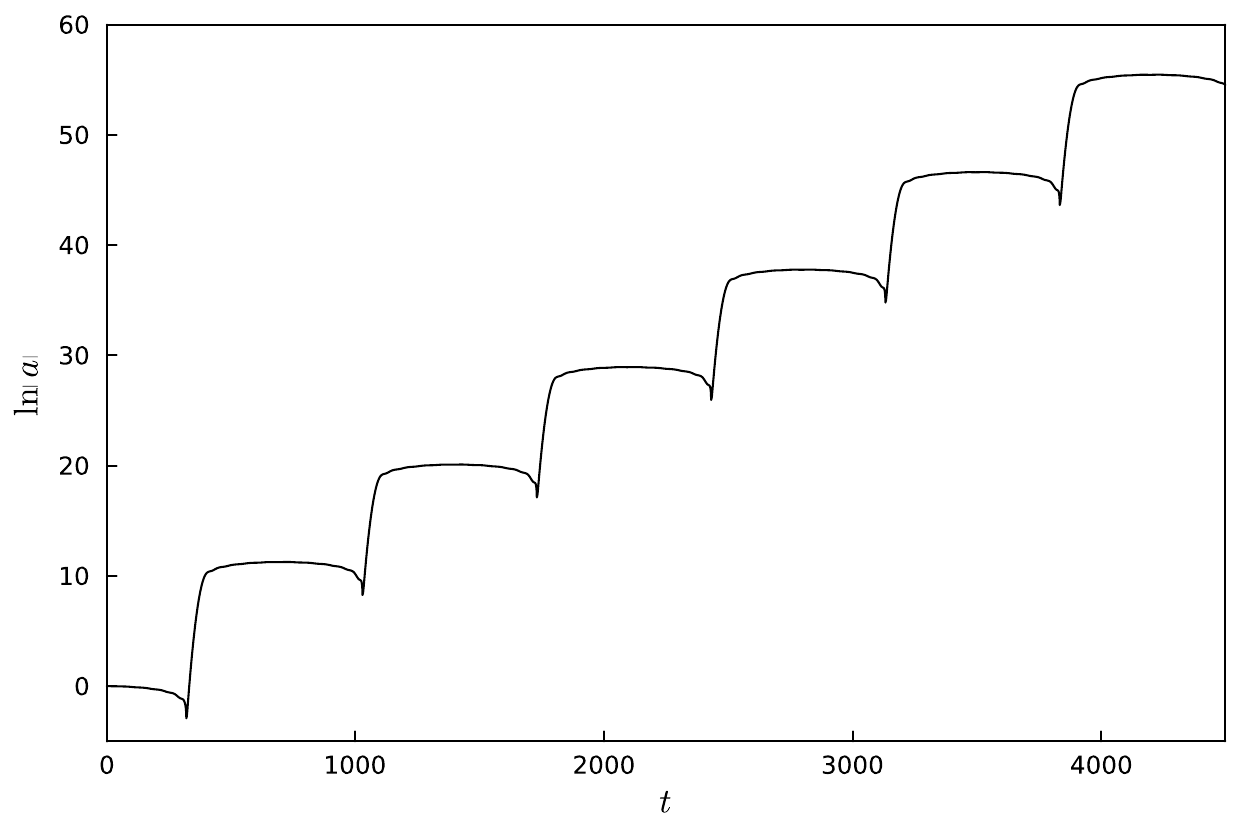}
\caption{The evolution of a cyclic Universe in the presence of a scalar field $\phi$ coupled nonminimally to curvature and a negative cosmological constant. The scale factor $a$ increases while both the amount of inflation and the lifetime of each cycle remain constant. The scalar field potential is $V(\phi)=V_0\phi^4+\Lambda$ with ~~$V_0=0.02$,  $\Lambda=-0.000018$. The coupling constant is $\xi=-1$. The bounce is set at the moment when ${\dot\phi}^2/2$ exceeds the bare reduced Planck mass to the fourth power. The initial conditions are ~~$\phi(0)=\dot\phi(0)=0.006$, and $H(0)$ is the negative root of the quadratic equation~(\ref{00}).}
\label{fig:lnavstscalar}
\end{figure}

    All asymptotic regimes for the system under investigation have been listed in~\cite{Carloni_2008}. In the minimally coupled scalar field case, the time arrow occurs since the expansion regime is potential-dominating ($\omega_\mathrm{eff}=-1$) while the contraction regime is the kinetic term dominated ($\omega_\mathrm{eff}=+1$). In the nonminimal coupling case, we use (\ref{rhopphi}) and define $\omega_\mathrm{eff}= p_{\phi}/\rho_{\phi}$. In principle, cross-terms in (\ref{rhopphi}) could be attributed to the geometry part, but not to energy and momentum of the scalar field, so there is a certain ambiguity in the definitions. Our choice has an advantage of using the same equation for the scale factor evolution --- it is clear from (\ref{00rho}) and (\ref{11p}) that the power index $n$ in the solution for the scale factor $a(t) \propto t^n$ is $n=2/(3\omega_\mathrm{eff}+3)$ as in the minimal coupling case. For the nonminimal case, an inflation is still possible, so in any case, the equation of state parameter is $\omega_\mathrm{eff}<-1/3$ at expansion for inflationary models. As to contraction, it is known that the kinetic regime as an attractor for a contracting Universe changes. In the model with nonminimal coupling with curvature, the single kinetic regime splits into two different regimes (see~\cite{we2,jarv2022global}), but only one of them is a past attractor. For the attractor we have $a(t)\propto t^{\alpha}$ (then $\dot H/H^2=-1/\alpha$) with 
\begin{equation}
\alpha=\frac{1}{3 - 12\xi - 2\sqrt{6\xi\left(6\xi - 1\right)}},
\end{equation}
which evidently leads to
\begin{equation}
\label{weff}
\omega_\mathrm{eff} = 1-8\xi-\frac{4}{3}\sqrt{6\xi(6\xi - 1)},
\end{equation}
where we use 
$$
\omega_\mathrm{eff}=\frac{p_{\phi}}{\rho_{\phi}}=-\frac{2\dot H+3H^2}{3H^2}=-\frac{2}{3}\frac{\dot H}{H^2}-1.
$$
The equation of state parameter is always larger than $1/3$ and is plotted as a function of the coupling constant $\xi$ in Fig.~\ref{fig:weffvsxi}. This means that for scalar field potentials allowing inflation we can expect a time asymmetry, as in the minimal coupling case. For the quartic potential this appears to be correct, as we have confirmed by numerical studies (see Fig.~\ref{fig:lnavstscalar}). This case has a particular importance since it corresponds to the Higgs inflation model if the coupling constant has observationally fixed value of $|\xi| = 17000$. Growing cycles of the scale factor appears clearly. We should note that, in the nonminimally coupled case, there is no unique preferable energy scale for the bounce since the effective Planck scale depends now on the value of the scalar field. Thus we can make the bounce either at the bare ($M_{Pl}=\sqrt{1/K}$) or the effective (${(M_{Pl})}_\mathrm{eff}=\sqrt{1/K-\xi B(\phi)}$) Planck scale. Both cases, however, lead to qualitatively the same picture. We also note that as the equation for the Hubble parameter~(\ref{00}) contains a cross-term linear in $H$, two roots of this equation have different signs and different absolute values $|H|$, so the bounce reverses the sign of $H$ but, unlike the minimal coupling case, does not keep its absolute value.
\begin{figure}[hbtp]
\centering
\includegraphics[scale=0.68]{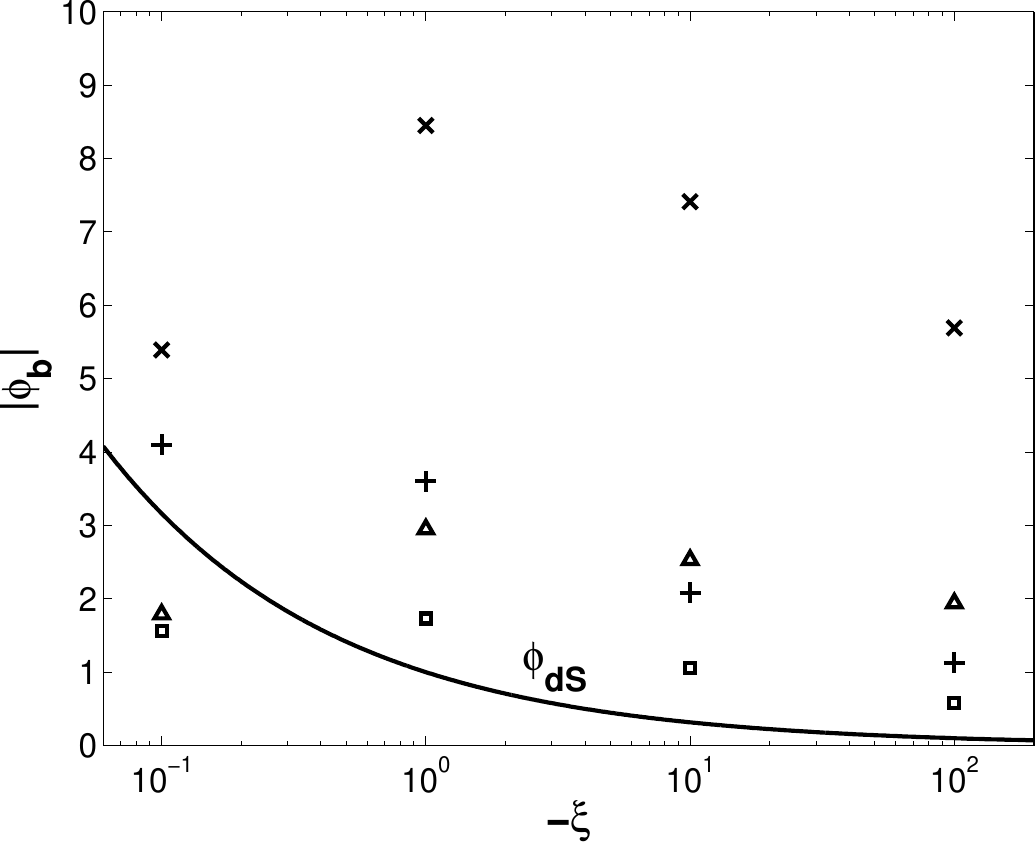}
\caption{The dependence of the critical value of the scalar field $|\phi_{b}|$ on the coupling constant~$-\xi$ for the nonminimal coupling model with $\xi<0$, the potential $V=V_0\phi^2$ with $V_0=0.01$. The critical value $|\phi_b|$ is found at the moment when ${\dot\phi}^2/2$ exceeds one of four different Plank masses to the fourth power: bare reduced (squares), bare non-reduced (pluses), effective reduced (triangles), effective non-reduced (crosses). The solid line is the de Sitter solution.}
\label{Fig3}
\end{figure}
\begin{figure}[!ht]
\centering
\includegraphics[scale=0.68]{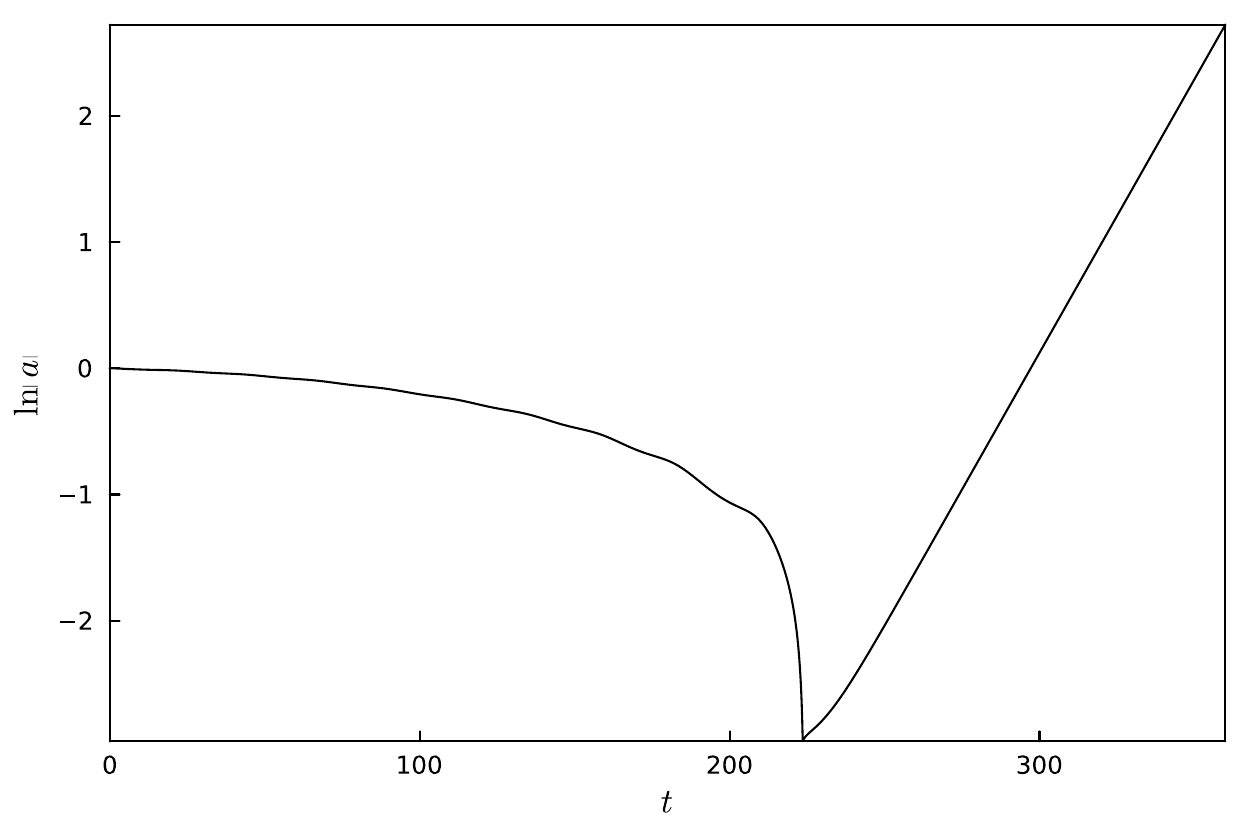}
\caption{The time dependence $\ln a(t)$ for the nonminimal coupling model with the potential $V(\phi)=~V_0\phi^2+\Lambda$, where $V_0=0.01$, $\Lambda=-0.00005$. The coupling constant is $\xi=-1$. The bounce is set at the moment when ${\dot\phi}^2/2$ exceeds the bare reduced Planck mass to the fourth power. The initial conditions are ~~$\phi(0)=\dot\phi(0)=0.01$, and ~~$H(0)$ is the negative root of the quadratic equation~(\ref{00}).}
\label{fig:lnavstscalarn2}
\end{figure}

    The quadratic potential represents, however, the case when the cycle scheme can fail, and the reason of the failure is {\it not} connected with the effective scalar field equation of state. Inability to close a cycle comes from the side of a turning point. It is known that a regime of infinite exponential growth of the scale factor is present for such model. It exists when the scalar field exceeds the value of the unstable de Sitter solution $\phi_{dS}=\pm\sqrt{-\frac{1}{K\xi}}$ existing for the quadratic potential. Most of the initial conditions with $\phi>\phi_{dS}$ lead to this regime of infinite scalar field and scale factor growth, 
\begin{equation}
\phi(t)\propto e^{\frac{2H_0\xi t}{4\xi-1}}, \quad a(t)\propto e^{H_0t},
\label{dS}
\end{equation}
where $H_0=const$, corresponding phase diagrams can be found in~\cite{we2,jarv2022global}. If the value of the scalar field at the bounce exceeds $\phi_{dS}$, the regime~(\ref{dS}) is triggered. In Fig.~\ref{Fig3} we present four possible choices for a bounce, with kinetic energy of the scalar field (which has the dimension of ${(\text{mass})}^4$) to be equal to either bare or effective Planck energy, which in its turn is defined via the reduced or nonreduced Planck mass (the latter is larger by a factor of $\sqrt{8 \pi}$). We can see that for large enough $|\xi|$ all four cases lead to the scalar field at bounce exceeding $\phi_{dS}$. Then, a rapid growth of the scale factor and the scalar field cannot be terminated by a positive spatial curvature or a small negative cosmological constant (see Fig.~\ref{fig:lnavstscalarn2}), and after one bounce the Universe enters eternal expansion. Moreover, since after the bounce the scalar field continues to grow, it can reach $\phi_{dS}$ even after the bounce with the same consequences. Our numerical integration shows, for example, that although the scalar field at bounce at $\xi=-0.1$ with the chosen value of $V_0=0.01$ is lower than $\phi_{dS}$, it reaches $\phi_{dS}$ after the bounce. If a bounce is made at the bare nonreduced Planck scale, only for $|\xi| < 0.063$ there emerges a cyclic behavior.

\section{Nonminimal kinetic coupling}
~~~~In this section we consider the situation where a cosmological hysteresis is totally absent. It can be realized in the theory of gravity with a scalar field possessing a nonminimal kinetic coupling to gravity proposed in~\cite{sushkov2009exact,PhysRevLett.105.011302}. In this model, the action is given by
\begin{equation}
\label{S2}
S =\int d^4x\sqrt{-g}\left (\frac{R}{K}-(g^{\mu\nu}+\kappa G^{\mu\nu})\partial_{\mu}\phi\partial_{\nu}\phi-2 V(\phi)\right),
\end{equation}
where the quantities ~$g=\mathrm{det}(g_{\mu\nu})$, ~$R$, ~$V(\phi)$,~ and ~$K$~ have the same meanings as in the previous section, $G_{\mu\nu}$ is the Einstein tensor, and $\kappa$ is a derivative coupling parameter with the dimensions of length squared.  

    This model belongs to the intensively studied type known as Horndeski models. The equations of motion in this type of model are second-order differential equations (so the nonminimal curvature coupling studied in the previous section also belongs to this class). Unlike the previous case, where the coupling function is, in general, not fixed, here the nonminimal term is unique --- only $G^{\mu\nu}$ gives second-order equations of motion. So this model is free from an arbitrary functions and can be studied extensively. For us here it is interesting because it contains a regime where the effective equation of state is the same for both expansion and contraction (see below).
\begin{figure}[!ht]
\centering
\includegraphics[scale=0.68]{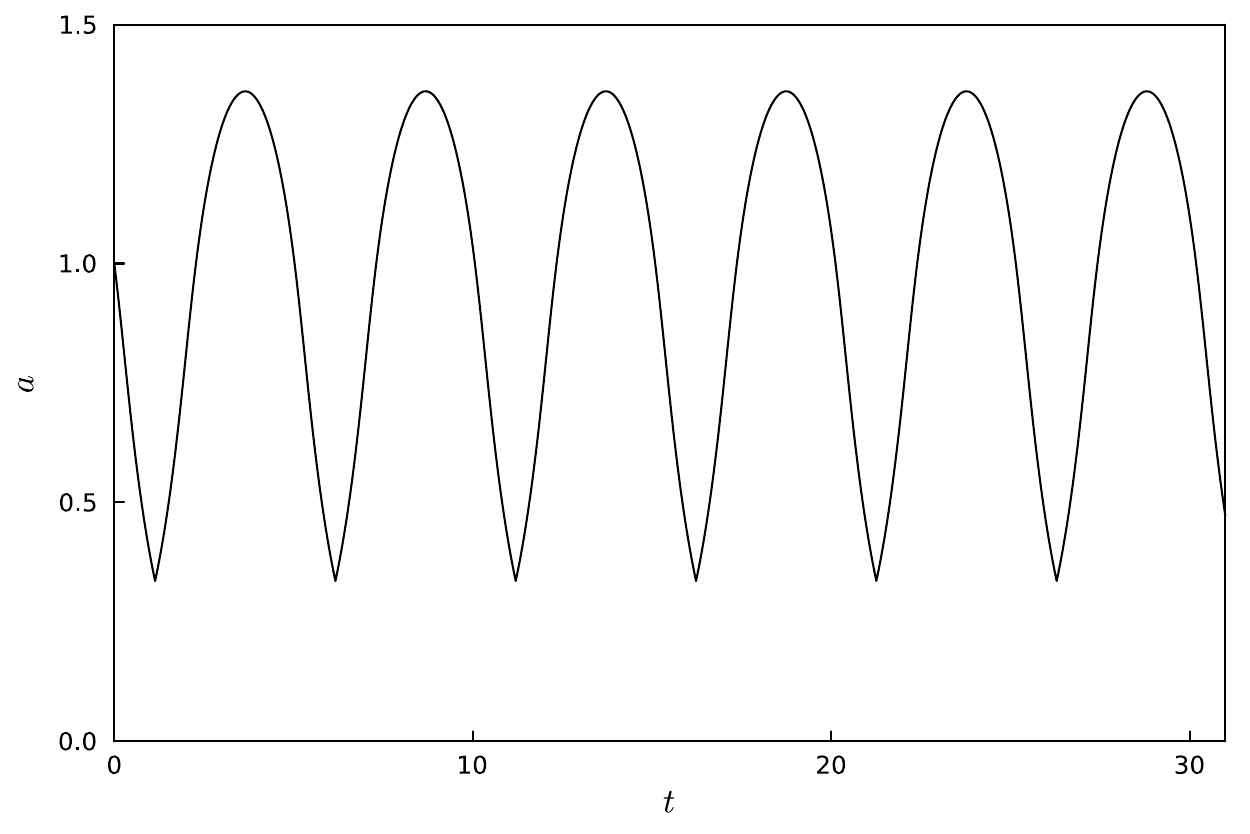}
\caption{The time dependence $a(t)$ for the nonminimal kinetic coupling model with the potential $V(\phi)=\Lambda=-0.01$. The coupling constant is $\kappa=0.1$. The initial conditions are ~~$\phi(0)=~\dot\phi(0)=~0.4$, ~~$a(0)=1$, and ~~$H(0)$ is the negative root of the quadratic equation~(\ref{00kin}).}
\label{fig:lnavstkinetic}
\end{figure}

    Variation of this action with respect to $(g^{\mu\nu})$ and $(\phi)$ gives the field equations
\begin{equation}
\label{eequakin}
G_{\mu \nu}=K \left[T_{\mu \nu}(\phi)+\kappa \Theta_{\mu\nu}(\phi) \right],
\end{equation}
\begin{equation}
\label{kgequakin}
\left[ g^{\mu\nu}+ \kappa G^{\mu \nu}\right]\nabla_{\mu}\nabla_{\nu} \phi=-V'(\phi),
\end{equation}
In the spatially flat FRW universe with the metric $ds^2=-dt^2+a^2(t)dl^2$, the time-time component of Eq.~(\ref{eequakin}) gives the energy constraint equation
\begin{equation}
\label{00kin}
3 H^2=K\left( \frac{1}{2}{\dot{\phi}}^2(1-9\kappa H^2)+V(\phi)\right) ,
\end{equation}
while the spatial diagonal components are 
\begin{equation}
\label{11kin}
2\dot H+3H^2=K\left[ -\frac{{\dot\phi}^2}{2}+V(\phi)-\kappa\left(\frac{{\dot\phi}^2}{2}(2\dot H+3H^2)+2H\dot\phi\ddot\phi\right)\right],
\end{equation}
and the Klein-Gordon equation reduces to
\begin{equation}
\label{KGkin}
\ddot\phi+3 H\dot\phi-3\kappa( H^2\ddot\phi+2 H\dot H\dot\phi+3 H^3\dot\phi)+V'(\phi)=0.
\end{equation}

    A remarkable feature of this model is that inflation can be realized even in the case
of a vanishing scalar field potential~\cite{sushkov2012realistic} if the coupling constant $\kappa$ is positive. What is also interesting, it is the inflation that represents in this case the past attractor for the corresponding cosmological evolution~\cite{we3}. This means that the equation of state for the scalar field at the contraction stage is $\omega_\mathrm{eff}=-1$. It is important for our goals now since it indicates that the effective equation of state during expansion is the same as at contraction. Thus we could expect that the time arrow we study in the present paper is absent for this particular case. A numerical simulation presented in Fig.~\ref{fig:lnavstkinetic} (we use a zero scalar field potential with a small negative cosmological constant here) confirms that.

\section{Conclusions}
~~~~In the present paper, we consider cyclic cosmological models in the presence of a nonminimally coupled scalar field. We check the  correspondence between contraction-expansion asymmetry of the effective equation of state of the scalar field and the property of the model to have an increasing cycles. This correspondence has been previously established for a minimally coupled scalar field. To achieve our goal, we consider two different models of nonminimal coupling, one is the curvature coupling, and the other is the kinetic coupling. It is known that while the above-mentioned asymmetry is present in the first case, the second model contains a regime where this asymmetry is absent. 

    Our results confirms this correspondence. The cycles grow in the curvature coupling model (where $\omega_\mathrm{eff}$ differs at expansion and contraction) and are of the same size in the kinetic coupling model (where $\omega_\mathrm{eff}$ is the same). We should, however, note that the cyclic behavior itself is absent in some cases of curvature coupling models. The reason is that a turnaround point can be absent. We have shown this for a quadratic scalar field potential. In this case, a small negative cosmological constant added to the model to ensure a turnaround point, does not work, and the expansion does not end. This means that we need a stronger method to get a recollaps in such a model. In principle, we can construct the turnaround ``by hand'', in the same way as we added the bounce. This is, however, questionable from the physical point of view since an artificial bounce is added in the large curvature regime, where we could expect corrections to the classical equations of motion, while a turnaround require modifications of the equations of motion at small curvature, which is less natural. On the other hand, when a cyclic behavior in the curvature coupling model is present, the cycles grow in time.

\section*{Acknowledgements}
~~~~The work of AT and MS is supported by RSF grant 21-12-00130. AT also thanks the Russian Government Program of Competitive Growth of Kazan Federal University. LA is supported by grants from the University of Iceland Research Fund. TV is supported by the Vice-rectorate for Research and Postgraduate B20131101 and CONCYTEC through the Research Teacher grant. AT is grateful to Lima University, where this work was initiated, for hospitality.

\end{document}